\documentclass[aps, pra, 
			   final, 
			   preprint]{revtex4-1}

\usepackage{amsmath}
\usepackage{amsfonts}
\usepackage{amssymb}
\usepackage{gensymb}
\usepackage{graphicx}
\usepackage[colorlinks=true, linkcolor=black, urlcolor=blue, citecolor=black, anchorcolor=black]{hyperref}

\graphicspath{{"../../../figures/"}}

\begin{document}
\title{Group and phase velocity mismatch fringes in triple sum-frequency spectroscopy}
\author{Darien J. Morrow}
\author{Daniel D. Kohler}
\author{John C. Wright}
	\email{wright@chem.wisc.edu}
\affiliation{Department of Chemistry,
	University of Wisconsin--Madison,
	1101 University Ave, 
	Madison, WI 53706, United States}

\date{\today}

\begin{abstract}
	The effects of group and phase velocity mismatch are well-known in optical harmonic generation, but the non-degenerate cases remain unexplored. 
	In this work we develop an analytic model which predicts velocity mismatch effects in non-degenerate triple sum-frequency mixing, TSF. 
	We verify this model experimentally using two tunable, ultrafast, short-wave-IR lasers to demonstrate spectral fringes in the TSF output from a 500 $\mu$m thick sapphire plate. 
	We find the spectral dependence of the TSF depends strongly on both the phase velocity and the group velocity differences between the input and output fields. 
	We define practical strategies for mitigating the impact of velocity mismatches.   
\end{abstract}

\maketitle

\section{Introduction}\label{S:Introduction}

Triple sum-frequency, TSF, generation is a multicolor four-wave mixing process in which the generated electric field has an output frequency defined by the sum of all three driving fields.
TSF is the four-wave mixing extension of sum-frequency generation (SFG), a three-wave mixing, ladder-climbing process, and the multicolor extension of third-harmonic generation (THG).  
In TSF, three electric fields drive an oscillating nonlinear polarization which generates the measured TSF field as defined by the medium's susceptibility.
This susceptibility is the sole source of analyte information. 
In TSF, the driving lasers' frequencies are scanned; when a driving field is resonant with a state, the susceptibility becomes large and the TSF intensity dramatically increases. 
Wright and coworkers are actively developing TSF as an analytical methodology which is sensitive to vibrational-electronic state coupling.\cite{Boyle_Wright_2013, Boyle_Wright_2013_001, Boyle_Wright_2014, Wright_2017}   

It is well known that spectroscopies which are defined by the sum of their driving frequencies and accomplished in normally dispersive samples cannot be phase-matched. 
This means that the emitted TSF field cannot maintain a cooperative phase relationship with the driven non-linear polarization for long distances because they travel with different velocities.\cite{Armstrong_Pershan_1962, Bloembergen_1965} 
Velocity mismatches cause the output to scale in a non-trivial way with sample length.
Ultrafast pulses further complicate the situation because the different fields travel with different group velocities and can temporally walk away from each other.\cite{Stoker_Keto_2005}
For instance, in THG microscopy, group velocity effects lead to an unusual depth dependence that can be mistaken for surface selectivity.\cite{Stoker_Keto_2005, Stoker_Keto_2006}
This non-trivial scaling between the non-linear polarization and the emitted field drastically complicates measurement of the susceptibility. 

As a rule of thumb, velocity-matching effects are mitigated by minimizing the excitation region's length, $L$, but the thinness required to satisfy this rule of thumb ($L<10 \:\mu\text{m}$ for the experiment explicated herein) can be structurally untenable.   
Structurally, thick windows or substrates are desirable for TSF spectroscopy (e.g. a thin film deposited on a thick substrate or a liquid sample sandwiched between two windows). 
In this work, we consider the response of a typical substrate in an ultrafast, non-resonant TSF experiment to demonstrate phase and group velocity effects.
We accomplish a 2-color TSF experiment with frequency $\omega_{\textrm{TSF}} = \omega_1 + 2\omega_2$ and spatial phase $\vec{k}_{\textrm{TSF}} = \vec{k}_1 + 2\vec{k}_2$.
We find that long samples and broadband excitation pulses lead to characteristic modulations in the output spectra defined by velocity matching conditions. 
These modulations depend on excitation color and will obscure the analyte response unless strategies are used to mitigate the observed fringes.
We define such strategies in the Discussion section.
Our formalism and findings easily extend to all wave-mixing processes whose output frequency is the sum of their input frequencies.

\section{Theory}\label{S:Theory}

In this section we solve the wave equation for TSF using pulsed excitation with finite bandwidth. 
Our derivation is informed by \textcite{Angerer_Sun_1999}'s frequency domain derivation of ultrafast SHG and Boyd's\cite{Boyd_2008} derivation of continuous wave (CW) SHG intensity.
Our derivation neglects the transverse evolution of the wave equation; these effects are important in experiments that tightly focus or have large beam crossing angles.\cite{Tasgal_Band_2004} 

The formation of the TSF electric field, $E_4(z,\omega)$, through a dispersive medium is given by Maxwell's scalar wave equation (presented in the frequency domain and in the SI unit system)\cite{Boyd_2008}:
\begin{equation}\label{E:wavew}
	\left[ \frac{\partial^2}{\partial z^2} + k^2(\omega)  \right] E_4  = \frac{\omega^2}{\epsilon_0 c^2} P_{\textrm{NL}}, 
\end{equation}
in which $\epsilon_0$ is the vacuum dielectric constant, $c$ is the vacuum speed of light, $P_{\textrm{NL}}$ is a non-linear polarization driven by the excitation fields, and $k(\omega)$ is the frequency-dependent spatial wavevector.  
For both the excitation pulses and the TSF output field, the spatial wavevector is described by a first-order Taylor expansion about the field's central frequency, $\omega_0$:
\begin{equation}\label{E:k}
\begin{split}
	k(\omega) &\approx k(\omega_0) + \left. \frac{\partial k}{\partial \omega} \right|_{\omega_0} (\omega - \omega_0) \\
			  &= v_p^{-1}\omega_0  + v_g^{-1} (\omega - \omega_0),
\end{split}
\end{equation}
where $v_p$ and $v_g$ are the phase and group velocity at $\omega_0$, respectively. 
The phase velocity is related to the refractive index, $n$, by $v_p = c/n$, and the group velocity is related to the phase velocity and refractive index by $v_g = v_p \left(1 + \frac{\omega}{n}\frac{\partial n}{\partial \omega}\right)$.
Truncating the Taylor series after the first-order neglects effects like group velocity dispersion. We also neglect effects like self-phase modulation. These effects are small because we work with sufficiently short samples and weak driving fields.\footnote{ 	We consider the effects of disregarding $\left. \frac{\partial^2 k}{\partial \omega^2} \right|_{\omega_0}$ in our Taylor expansion by comparing the accrued pulse duration (chirp), $\delta t \approx \left. \frac{\partial^2 k}{\partial^2 \omega} \right|_{\omega_0} \Delta  \omega \cdot L$, to the original pulse duration, $\Delta t \approx \frac{0.441 \cdot 2}{\Delta \omega}$. For our sample and experimental conditions we find $\frac{\delta t}{\Delta t} \approx 0.1 \%$. Additionally, we may consider the prominence of self-phase modulation for our experimental conditions. \textcite{Siegman_1986} notes that the length scale, $L$, over which self-phase modulation is significant goes as $L = \frac{\lambda}{2\pi I n_{2I}}$ where $I$ is the intensity at the beam waist and $n_{2I}$ is the nonlinear refractive index. Using our experimental parameters and \textcite{Major_Smith_2004}'s value of $n_{2I} = 3\times10^{-16} \; \textrm{cm}^2/\textrm{W}$ for $\lambda=1300 \; \textrm{nm}$, we find that $L\approx0.3 \; \textrm{cm}$, which is an order of magnitude longer than our sample. Given both of these calculations, second-order effects are much smaller than first-order effects given our pulse bandwidth and sample length. We think of the intensity of driving fields at which our work is accomplished as being sufficient to see generated third-order response against black, but not sufficient to observe third-order effects (self-phase modulation) in the driving fields.}

Our electric fields have Gaussian envelopes:
\begin{equation} \label{E:field}
	E_j(z, \omega ; \omega_j) = A_j(z) \exp{\left[ik_j(\omega)z\right]}\exp{\left[-\frac{(\omega-\omega_j)^2}{2\sigma_j^2}\right]}  ,
\end{equation}
where $\omega_j$ is the spectral center, $\sigma_j$ is the spectral bandwidth, $k_j(\omega)$ is $k(\omega)$ expanded such that $\omega_0 = \omega_j$,  and $A_j(z)$ is the amplitude through the sample.  As we shall show, an analytic solution to the wave equation results when we assume this form for both the driving excitation fields ($E_1$, $E_2$, and $E_3$) as well as the generated TSF field ($E_4$).
Using this definition for our TSF field, and invoking the slowly varying envelope approximation $\left( \left| \frac{\partial^2 A_4(z) }{\partial z^2} \right| \ll \left| 2ik_4(\omega)\frac{\partial A_4(z)}{\partial z} \right| \right)$ in order to disregard the second order derivative, Eqn. \ref{E:wavew} becomes
\begin{equation} \label{E:before_SVEA}
\frac{\partial A_4(z) }{\partial z}  \exp{\left[ik_4(\omega)z\right]}\exp{\left[-\frac{(\omega-\omega_4)^2}{2\sigma_4^2}\right]} = \frac{\omega^2}{2ik_4(\omega) \epsilon_0 c^2}P_{\mathrm{NL}}(z,\omega).
\end{equation} 

We now consider the form of the non-linear polarization.  In the convention of \textcite{Maker_Terhune_1965}, $P_\text{NL}$ is given by
\begin{equation}\label{E:polar_mt} 
	P_{\text{NL}}(z,\omega_4=\omega_1 + \omega_2 + \omega_3) = \epsilon_0 \chi^{(3)}\left(\omega_4; \omega_1, \omega_2, \omega_3 \right) E_1(\omega_1) E_2(\omega_2) E_3(\omega_3)
\end{equation}
in which $\chi^{(3)}$ is the third-order susceptibility (we suppress all tensors in this derivation).
Equation \ref{E:polar_mt} neglects the buildup of polarization that can occur with resonant, impulsive excitation.\cite{Kohler_Wright_2017} 
It is applicable for this work since transparent materials lack visible and near-IR resonances. 

To account for our finite pulse bandwidth, the non-linear polarization is the weighted average of all incident field components:
\begin{equation} 
\begin{split}\label{E:polar_int}
	P_{\text{NL}}(z,\omega_4) =& \epsilon_0 \iiint\limits_{-\infty}^{+\infty}
	\chi^{(3)}\left(\omega_\alpha + \omega_\beta + \omega_\gamma; \omega_\alpha, \omega_\beta, \omega_\gamma, \right) \\
	& \qquad \times E_1(z, \omega_\alpha ; \omega_1) E_2(z, \omega_\beta ; \omega_2) E_3(z, \omega_\gamma ; \omega_3) \\
	& \qquad \times \delta \left(\omega_4 -  \omega_\alpha - \omega_\beta - \omega_\gamma\right) \textrm{d}\omega_\alpha \textrm{d}\omega_\beta \textrm{d}\omega_\gamma,
\end{split}
\end{equation}
in which $\delta$ is the Dirac delta distribution.  Furthermore, since transparent materials have a pseudo-flat spectral response, $\chi^{(3)}$ is well-approximated as a constant and may be removed from the integral.  

For this work, we assume all driving fields have the same bandwidth, $\sigma_j=\sigma$.
Assuming $\chi^{(3)}$ is constant and approximating $k(\omega)$ with a first-order Taylor expansion, Eqn. \ref{E:polar_int} can be evaluated\footnote{
	We use the well-known relation $\int_{-\infty}^{\infty} \exp{\left[-\left(ax^2+bx\right)\right]}\textrm{d}x = \sqrt{\frac{\pi}{a}}\exp{\left[\frac{b^2}{4a}\right]}$ to integrate Eqn. \ref{E:polar}. 
} as: 
\begin{equation} \label{E:polar}
\begin{split}
P_{\mathrm{NL}}(z,\omega) =&  P(z) \exp{\left[-\frac{\left(\omega -\omega_1 -\omega_2 -\omega_3  \right)^2}{6\sigma^2}\right]} \\
	&\times \exp{\left[iz\left(v_{p,1}^{-1}\omega_1 + v_{p,2}^{-1}\omega_2 + v_{p,3}^{-1}\omega_3 \right)\right]} \\
	&\times \exp{\left[\frac{iz}{3}\left(v_{g,1}^{-1}+v_{g,2}^{-1}+v_{g,3}^{-1}\right)\left(\omega -\omega_1 -\omega_2 -\omega_3\right)\right]} \\
	&\times \exp{\left[\frac{\sigma^2 z^2}{3}\left(v_{g,1}^{-1}v_{g,2}^{-1}+v_{g,1}^{-1}v_{g,3}^{-1}+ v_{g,2}^{-1}v_{g,3}^{-1} -v_{g,1}^{-2}-v_{g,2}^{-2}-v_{g,3}^{-2} \right)\right]} 
\end{split}
\end{equation}  
in which we defined the spatial amplitude term
\begin{equation}
	P(z) \equiv \frac{2\pi \epsilon_0 \sigma^2}{\sqrt{3}} \chi^{(3)} A_1(z) A_2(z) A_3(z).
\end{equation}
Inspection of Eqn. \ref{E:polar} shows that the spectral bandwidth of the polarization is $\sqrt{3}$ larger than the driving fields and centered at the TSF frequency $\omega = \omega_1+\omega_2+\omega_3$.  
The last multiplier in Eqn. \ref{E:polar} depends on $\sigma^2 z^2$ and the differences in group velocity among the driving pulses. By the Cauchy-Schwarz inequality, the exponent is always negative so that the multiplier is bounded to $(0,1]$. 
This term captures how group velocity differences cause the pulsed excitation beams to temporally walk off from each other. 
Such effects may become important in TSF using disparate driving frequencies.  
For simplicity, we approximate this term as unity, which is valid when all driving fields have sufficiently similar group velocities. 
Since we have assumed a non-resonant medium with a shallow focus, there is no depletion of the excitation fields so we approximate the amplitudes as constant:  $P(z)=P$.

We now substitute Eqn. \ref{E:polar} into Eqn. \ref{E:before_SVEA}.
We assume the generated TSF field has the same spectral properties as $P_\text{NL}$ ($\sigma_4=\sqrt{3} \sigma$ and $\omega_4 = \omega_1 + \omega_2 + \omega_3$), and so Eqn. \ref{E:before_SVEA} simplifies to:
\begin{equation}\label{E:wavew2}
	\frac{\partial A_4(z) }{\partial z} = P\frac{\omega^2}{2ik_4(\omega)\epsilon_0c^2} \exp{\left[iz\left(\Delta k + \left(\omega-\omega_4\right)\Delta v_{g}^{-1}\right)\right]},
\end{equation} 
where the phase velocity mismatch, $\Delta k$, and the group velocity mismatch, $\Delta v_{g}$, are defined according to:
\begin{align}
	\Delta k 		  &\equiv v_{p,1}^{-1}\omega_1 + v_{p,2}^{-1}\omega_2 + v_{p,3}^{-1}\omega_3 - v_{p,4}^{-1}\omega_4, \\
	\Delta v_{g}^{-1} &\equiv \frac{v_{g,1}^{-1}+v_{g,2}^{-1}+v_{g,3}^{-1}}{3} - v_{g,4}^{-1} .
\end{align}
Integration of Eqn. \ref{E:wavew2} (from $z=0$ to $L$) yields
\begin{align} \label{E:A4}
A_4(L, \omega) = \frac{P\omega^2}{2k_4(\omega)\epsilon_0c^2} \left( \frac{1-\exp{\left[iL\left(\Delta k + \left(\omega-\omega_4\right)\Delta v_{g}^{-1}\right)\right]}}{\Delta k + \left(\omega-\omega_4\right)\Delta v_{g}^{-1}}\right),
\end{align} 
where $L$ is the sample path length.  
The total TSF field is then
\begin{equation}\label{E:E4}
E_4(L, \omega; \omega_4) = A_4(L, \omega)\exp{\left[ik_4(\omega)L\right]}\exp{\left[-\frac{(\omega-\omega_4)^2}{6\sigma^2}\right]}.
\end{equation}
For intensity-level detection, $I_4 \propto |E_4|^2$, we arrive at the key equation for this work:
\begin{equation} \label{E:main}
I_4(L, \omega) \propto I_1 I_2 I_3 L^2 \mathrm{sinc}^2\left[\frac{\left(\Delta k + \left(\omega-\omega_4\right)\Delta v_{g}^{-1}\right)L}{2}\right] \exp{\left[-\frac{(\omega-\omega_4)^2}{3\sigma^2}\right]}.
\end{equation}
Equation \ref{E:main} describes the expected TSF signal as $L$, $\Delta k$, and $\Delta v_{g}^{-1}$ change. 
If there is no group velocity mismatch, $\Delta v_{g}^{-1} = 0$, or if we measure at the center of the output pulse, $\omega=\omega_4$, then we recover the same $\Delta k L$ periodicity known in SHG and other processes.\cite{Boyd_2008, Shen_1984, Maker_Savage_1962}
Fringes defined by $\Delta k L$ will hereafter be called \emph{phase mismatch fringes}. 
These are also referred to as Maker-Terhune-type oscillations.\cite{Maker_Savage_1962, Mlejnek_Bloembergen_1999}
We note the expected intensity is periodically dependent on $\left(\Delta k + \left(\omega-\omega_4\right)\Delta v_{g}^{-1}\right)L$.
Normally, minimizing $\Delta k L$ maximizes the output intensity, but for ultrafast pulses, the group velocity mismatch is also important. 
The  $\left(\left(\omega-\omega_4\right)\Delta v_{g}^{-1}\right)L$ term will result in periodicities of the spectrally resolved output for a given color combination of pulses (hereafter called \emph{group mismatch fringes}).
This spectral dependency on group velocity mismatch is known for SHG. \cite{Trabs_Petrov_2015, Rassoul_Hache_1997, Glenn_1969, Sidick_Dienes_1995, Diels_Rudolph_1996}

\begin{figure}
	\includegraphics[width=0.5 \textwidth]{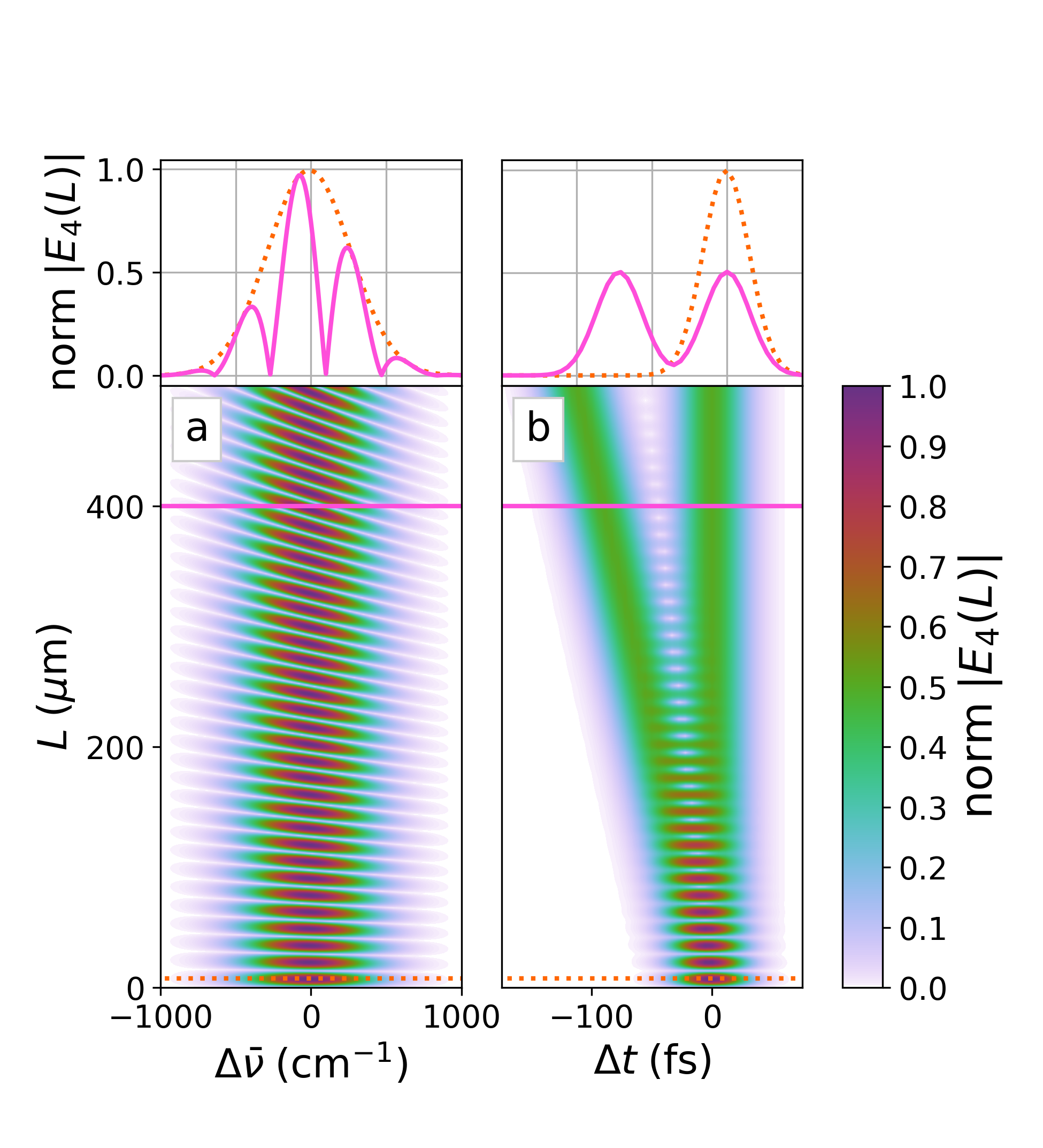}
	\caption{\label{F:degenerate}
		(Color online) Pulsed excitation model for degenerate excitation ($\bar{\nu}_1 = \bar{\nu}_2 = \bar{\nu}_3 = 7700 \ \text{cm}^-1$)  in sapphire with sample lengths up to $500 \ \mu \text{m}$. 
		(a) Frequency distribution of the output against sapphire sample length.  
		The frequency axis is referenced to the TSF frequency center, $\omega_4 = \omega_1 + \omega_2 + \omega_3$.
		(b) The temporal envelope of the output against sapphire sample length.  
		The time axis is referenced relative to when a TSF pulse which was generated at the front of the sample leaves the sample.  
		In both plots, the small plots overhead show frequency/time cross-sections of $L=10 \ \mu \text{m}$ (dotted orange) and $L=400 \ \mu \text{m}$ (solid dark pink).  
		The driving fields have width $\sigma = 160 \ \text{cm}^{-1}$.
	}
\end{figure}

To understand the consequences of this model, we calculate the electric field generated through a sapphire substrate as a function of sample length, $L$. 
We use the refractive index of sapphire as measured by \textcite{Malitson_1962}; for the range of excitation frequencies we survey, $\left|2\pi / \Delta k\right| \sim 15 \ \mu \text{m}$ and $\left|2\pi  \Delta v_g\right| \sim 30 \ \mu \text{m} / \text{fs}$. 
Figure \ref{F:degenerate} shows the calculated TSF field that results over a range of different substrate lengths. 
By showing the range of substrate lengths, one can observe the ``build-up'' of TSF through the sample.  
The frequency-domain (Fig. \ref{F:degenerate}a) and time-domain (Fig. \ref{F:degenerate}b) representations of the TSF field provide different insights on the propagation. We explore both representations to give a thorough picture of the propagation effects.  

As sample length increases, higher-order propagation effects are needed to explain the output.  
For the shortest path lengths ($L \ll \Delta k^{-1}$), phase mismatch and group velocity mismatch do not strongly influence the output and signal grows quadratically with $L$.  
Between the shortest path lengths and $\sim 50 \mu \text{m}$, ($\Delta k^{-1} < L < \Delta v_g \Delta_t$), signal output modulates with phase mismatch fringes.  
The modulation only depends on the sample length.
If CW driving lasers were in use, we would \emph{only} see these phase mismatch fringes.

At path lengths longer than $\sim 50 \ \mu \text{m}$, the pulsed nature of the propagation becomes essential to explain the evolution. 
In the frequency domain, these path lengths are large enough to resolve periodicities across the bandwidth of the TSF output.
The fringes, which were horizontal at smaller path lengths, now accrue a tilt that gives them a mixed frequency/path length dependence. 
The accrued tilt is defined by the color dependence in Eqn. \ref{E:main}, which gives modulations in the frequency distribution. 
In the time domain (Fig. \ref{F:degenerate}b), the group velocity difference is large enough that the driving field walks off of the initial TSF polarization created at the front of the sample.  
In effect, this walkoff causes the Gaussian pulse to break into two distinct pulses separated by time $\Delta v_g L$.  
The delay corresponds to the TSF field created at the back of the sample exiting the sample sooner than the field created at the front of the sample.
There is no TSF field in between the pulses because of symmetric, destructive interference of the phase mismatch fringes as previously seen in THG microscopy.\cite{Stoker_Keto_2005, Stoker_Keto_2006} 
Only electric fields generated at the sample edges contribute significantly to the observed output---electric fields generated at different planes in the sample interior are out of phase with each other and thus destructively interfere.\cite{Stoker_Keto_2005}   
Others have observed and explained this type of separation in SHG. \cite{Manassah_1988, Noordam_vanLindenvandenHeuvell_1990, Mlejnek_Bloembergen_1999}

\section{Experimental}\label{S:Experimental}

An ultrafast oscillator (Spectra-Physics, Tsunami) seeds a regenerative amplifier (Spectra-Physics, Spitfire Pro XP) which creates ultrafast pulses ($\sim 35\; \text{fs}$) centered at 12500 cm\textsuperscript{-1} with a 1 kHz repetition rate. 
These pulses pump two optical parametric amplifiers, OPAs, (Light Conversion, TOPAS-C) which we label ``OPA1'' and ``OPA2''. 
The OPAs are operated in the `signal' region for this experiment and their motors are tuned to maximize the smoothness of the OPA's tuning curve (see Fig. \ref{F:OPAs}a-d for OPA power curves and tuning tests). 
Their output, which we label $\omega_1$ and $\omega_2$, ranges from 6200 to 8700 cm\textsuperscript{-1}. 
A silicon wafer (0.4 mm thick) acts as a low-pass filter (cutoff: $\sim$8900 cm\textsuperscript{-1}) for removal of residual 12500 cm\textsuperscript{-1} pump light. 
A motorized (Newport, MFA-CC) retro-reflector defines the time delay, $\tau_{21}$, between the two pulses.
The relative delay of different colors of light caused by dispersion of transmissive optics is actively corrected by offsetting the $\tau_{21}$ set-point for each possible color combination.
The offset is empirically defined by maximizing transmitted TSF signal---see Fig. \ref{F:OPAs}e,f for the measured offset. 
A spherical mirror ($f=1$ m) focuses the two beams onto the sample (500 $\mu$m thick, double side polished sapphire) with each beam being 1\degree\space from surface normal (2\degree\space between beams).
The width of the Gaussian mode at the sample position is  $\sim 375\; \mu\text{m}$; incident pulse energies are $\sim 10\; \mu \text{J}$ ($\omega_2$) and $\sim 1\; \mu \text{J}$ ($\omega_1$) per pulse. 
The transmitted, spatially and temporally coherent output from the sample is spatially isolated in the $k_1 + 2k_2$ direction with an aperture, focused into a monochromator (HORIBA Jobin Yvon MicroHR, 140 mm focal length, with a 1200 nm blaze and 150 grooves per mm grating), and homodyne-detected (intensity level) with a thermoelectrically cooled PMT (Hamamatsu Photonics, H7422-20).
This PMT has a responsivity which changes by a factor of $\sim$4 over the range of detected light. 
All collected TSF spectra are shown on the amplitude level (in post-processing we take the square root of the detected/recorded intensity).
The acquisition software which controls all motors and records data is open source, written in Python, and available at \url{http://github.com/wright-group/PyCMDS}.
The Python computing language and the NumPy, SciPy, and Matplotlib libraries were used to collect, analyze, and represent the data presented in this work.\cite{vanRossum_2001, Jones_2001, vanderWalt_Varoquaux_2011, Hunter_2007} 

\section{Results}\label{S:Results}

\begin{figure*}[!htbp]
	\includegraphics[width=\textwidth]{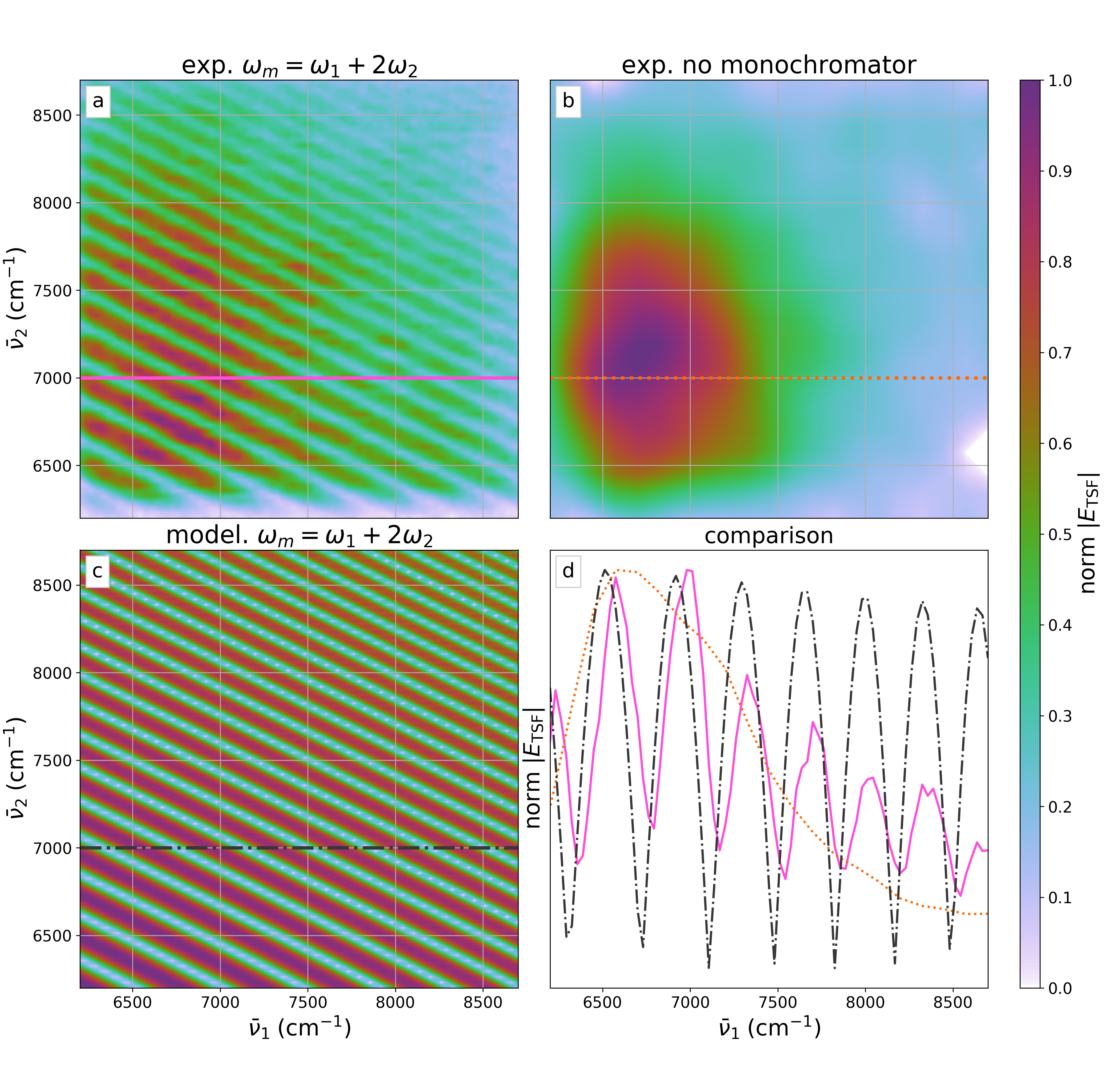}
	\caption{\label{F:w1w2}(Color online) TSF amplitude at multiple combinations of pump colors---a juxtaposition between experiment and model. Experimental spectra (a and b) are represented as the square-root of the detected intensity. These spectra go to zero near the edges due to a lack of driving laser intensity---see powercurves in SI. We note that (b) has been lightly smoothed. (c) is our model's prediction assuming $\omega=\omega_1+2\omega_2$ with effectively no spectral bandwidth of resolution. (d) compares a color and linestyle coded trace from each of (a, b, and c).}
\end{figure*}

We have described and shown the oscillatory nature of the TSF output as a function of sample length. 
However, when using TSF as an analytical method, the sample/substrate length is generally constant while the carrier frequencies of the driving pulses are scanned across resonances.
This scanning of carrier frequencies changes $\Delta k$ and $\Delta v_{g}^{-1}$ and can cause velocity mismatch fringes.
In order to observe these effects we performed a TSF experiment using two tunable, ultrafast pulses where $\omega_{\textrm{TSF}} = \omega_1 + 2\omega_2$ and $\vec{k}_{\textrm{TSF}} = \vec{k}_1 + 2\vec{k}_2$. 
The pulses have a bandwidth of $\sigma \approx 160\; \mathrm{cm}^{-1}$. 
Figure \ref{F:w1w2} shows the normalized TSF magnitude as a function of the two excitation frequencies. 
Figures \ref{F:w1w2}a,b show the experimental data with and without a tracking monochromator ($\omega=\omega_{\text{measured}}=\omega_1 + 2\omega_2$), respectively.\footnote{The spectra shown in Fig. \ref{F:w1w2}b was acquired with our monochromator/grating in ``0th order mode'' which effectively passed all colors to the detector as if it were a lossy mirror.} 
Figure \ref{F:w1w2}a displays deep periodicities along both axes. 
We are able to reproduce these periodicities with our model---see Fig. \ref{F:w1w2}c.
With a tracking monochromator, all periodicities are exclusively due to the changing phase velocity mismatch, $\Delta k$, between the the TSF emission and polarization.
Without a monochromator (Fig. \ref{F:w1w2}b), there are no fringes. We observe a peaked spectral profile which roughly follows the intensity profiles of our excitation lasers and detector spectral response function (much the same as the envelope of Fig. \ref{F:w1w2}a). 
In other words, for any combination of $\omega_1$ and $\omega_2$, TSF amplitude is created; however, the central frequency may not have appreciable amplitude due to phase mismatch effects. 

\begin{figure}[!htbp]
	\includegraphics[width=.5\textwidth]{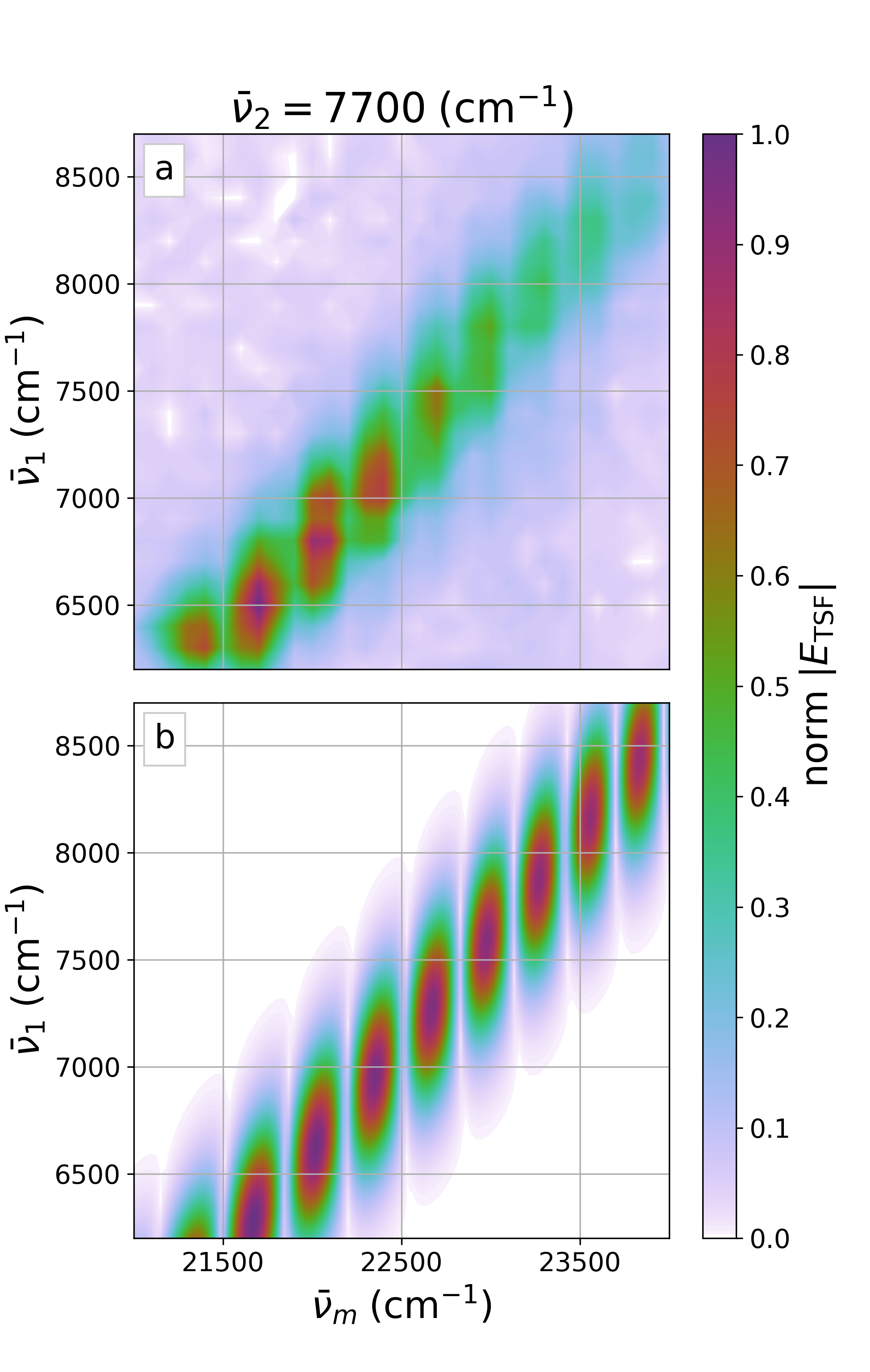}
	\caption{\label{F:mono}(Color online) TSF amplitude for multiple combinations of pump and monochromator color. The experimental spectrum (a) is represented as the square-root of the detected intensity. Subplot (a) shows experiment while subplot (b) shows our model's prediction.}
\end{figure}

In order to clarify the monochromator's role in the observation of spectral fringes, we scanned both $\omega_1$ and $\omega_m$ for a set $\omega_2$ frequency. 
The results are shown in Fig. \ref{F:mono}a. 
The total signal lies along the line $\omega_m - \omega_1 = \textrm{constant}$, but modulations are present along this line. 
These modulations are the same as those observed in Fig. \ref{F:w1w2}a. 
Figure \ref{F:mono}b shows the TSF amplitude as predicted by our model.
All periodicities along the $\omega_m$ axis are due to the $(\omega-\omega_4)\Delta v_{g}^{-1}$ term in Eqn. \ref{E:A4} (group mismatch fringes). 
We note that there is a slight curvature in the periodicities as $\omega_1$ changes; this curvature is due to the changes in group velocity of $\omega_1$ and $\omega_4$ not perfectly offsetting each other as $\omega_1$ changes.
Sapphire has fairly static group velocity differences between the excitation and emission frequencies explored in this work, so the curvature is slight.

\clearpage 

\section{Discussion}\label{S:Discussion}

Transparent materials are foundational components in optical sample cells because they are inactive as absorbers over spectral regions of interest. 
However, these materials do have substantial refractive indices. 
Consequently, they are bright in many non-linear experiments that are sensitive to both absorption and refraction and thus form a background signal that must be taken into account.
By exploring the multidimensional TSF spectrum of sapphire, we have shown that TSF spectroscopy can have complex and significant backgrounds from transparent materials used as windows or substrates.  
Unlike window contributions in other non-linear spectroscopies (cf. \textcite{Murdoch_Wright_2000}), window/substrate contributions to TSF are highly modulated in their output amplitude.  
These modulations can obscure analyte line shapes, especially when the modulation periodicity ($\Delta v_g^{-1}$) is comparable to the bandwidth of analyte features.  

These potential complications can be avoided in a variety of ways.  
The most direct approach is to keep material path lengths short ($L < \frac{2\pi}{\Delta k + (\omega-\omega_4)\Delta v_g^{-1}}$), which prevents the formation of mismatch fringes entirely. 	
This path length criterion is a modification of the CW standard of using samples thinner than $2\pi / |\Delta k |$.
Figure \ref{F:degenerate} shows that sapphire samples and substrates thinner than $\sim 10 \ \mu\text{m}$ fall within this standard.
Additionally, for the ranges of frequencies explored in this experiment, $\Delta k \gg \sqrt{3}\sigma \Delta v_g^{-1}$ (greater by a factor of $>35$) so just as in the CW case, $\Delta k$ defines the critical dependence on length that the experimentalist must consider for these thin samples.

Sufficiently short material path lengths are often impractical because they are structurally weak.  
If thick sample cells are required, a reflective geometry can mitigate background effects. 
Reflected (epi) THG has an effective penetration depth of  $\sim \lambda_{\textrm{fundamental}} / 12\pi$ which is $\sim 40 \ \text{nm}$ for our experiments. 
This small interaction distance is within the ``thin sample'' limit and therefore is not affected by mismatch effects. The small depth also keeps the amplitude of the background much smaller than the asymptotic limit, shown in Fig. \ref{F:degenerate}.
In a similar vein, researchers doing coherent anti-Stokes Raman and transient grating spectroscopies have used reflective geometries to efficiently discriminate against background signal.\cite{Volkmer_Xie_2001, Czech_Wright_2015}  
We also note that some groups have already accomplished THG in a reflective geometry in order to mitigate absorptive losses and focusing effects.\cite{Wang_Baardsen_1969, Bey_Rabin_1968}

For another option, we note that mismatch fringes are only observed if the output field is spectrally resolved: if the spectral (angular frequency) resolution, $R$, is worse than $R \approx L \Delta v_g^{-1}$, then the fringes are washed out.  
Decreasing the resolution of a monochromator, using no monochromator (as in Fig. \ref{F:w1w2}b), or using sufficiently \textit{long} material path lengths (large $L$) can all remove velocity mismatch effects. 
These are effectively a smoothing of Eqn. \ref{E:main} with respect to $\omega$ because the monochromator is incapable of resolving the fast oscillations.
Importantly, the decrease of output resolution may not affect the instrumental resolution as it pertains to $\chi^{(3)}$, because the bandwidth of the excitation fields already broadens the resolution.\cite{Kohler_Wright_2017}  

In light of our understanding of the TSF generation in non-resonant media, it is prudent to consider how resonant analytes will affect pulse propagation.
Unlike the window/substrate materials we have studied here, input frequencies will be scanned about analyte resonances, which can introduce dramatic pulse distortions that require a higher-order (and complex-valued) expansion of Eqn. \ref{E:k}.  
This potentially makes analyte TSF polarizations much different from the normally dispersive case analyzed here, because dispersion can be anomalous and large, and absorption is strong.\cite{Crisp_1970}
These complications are avoidable in cases of small analyte loading.
We note that it is common practice to keep analyte loading small enough ($\text{OD} < 0.3$) to avoid depletion of the pulse fields and the consequent spectral\cite{Carlson_Wright_1989, Kornau_Wright_2011} and temporal\cite{Kinrot_Prior_1994} signal distortions.

\section{Conclusion}\label{S:Conclusion}

The use of tunable ultrafast excitation pulses in triple sum frequency spectroscopy requires extension of previous treatments of phase matching effects to include group velocity mismatches. 
The group velocity mismatch fringes appear as both periodic modulations in the frequency distribution of the output or changing temporal delays between the output beams created near the front and back surfaces of the sample. 
If a monochromator is used to isolate the triple sum frequency signal, there will be interference effects between the beams. 
These effects create fringes that are defined by $\left|\sin{\left(\frac{\left(\Delta k + (\omega-\omega_4)\Delta v_g^{-1}\right) L}{2}\right)}\right|$ where the first and second terms of the argument describe the wave vector (phase velocity) mismatch and group velocity mismatch, respectively. 
The fringes create modulations in multidimensional triple sum frequency spectra as the excitation or monochromator frequencies are scanned. 
The modulations can complicate and obscure spectral features in samples containing resonances. 
These effects can be minimized by using short samples or keeping the output resolution at or lower than the pulse bandwidth.

\section*{Supplementary Material}
	All data and the workup/representation/simulation script are available for download at \url{http://dx.doi.org/10.17605/osf.io/emgta}. 
	
\begin{acknowledgments}
	We acknowledge support from the Department of Energy, Office of Basic Energy Sciences, Division of Materials Sciences and Engineering, under award DE-FG02-09ER46664.	
\end{acknowledgments}

\appendix

\section{Calculation of phase and group velocities}

\begin{figure}[!htbp]
	\includegraphics[width=.5\textwidth]{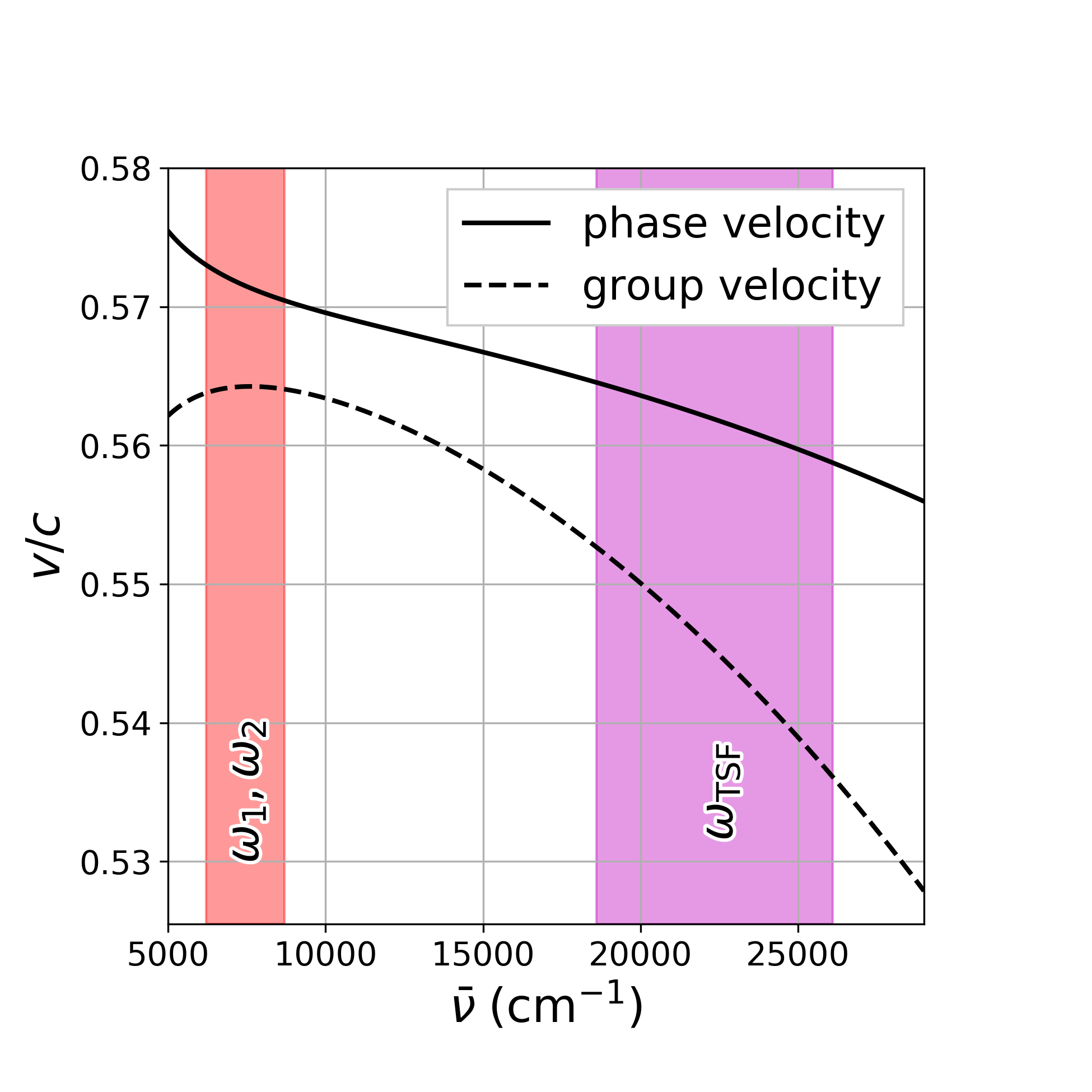}
	\caption{\label{F:velocities}(Color online)  Phase and group velocities ($v_p$ and $v_g$, respectively) in sapphire. Calculated from refractive index data measured by \textcite{Malitson_1962}. The salmon and magenta colored regions represent the experimental colors explored in this work by our pump lasers (labeled $\omega_1, \omega_2$) and TSF output (labeled $\omega_{\textrm{TSF}}$), respectively.}
\end{figure}

We use the following relations to calculate the phase and group velocity from refractive index data: 
\begin{align}
v_p(\omega) &= \frac{c}{n(\omega)} \\
v_g(\omega) &= \frac{c}{n(\omega) + \omega \frac{\partial n}{\partial \omega}} .
\end{align}
We use the refractive index of sapphire as measured by \textcite{Malitson_1962}. Our results are shown in Fig. \ref{F:velocities}.

\section{OPA output characterization and correction}

\begin{figure*}[!htbp]
	\includegraphics[width=1\textwidth]{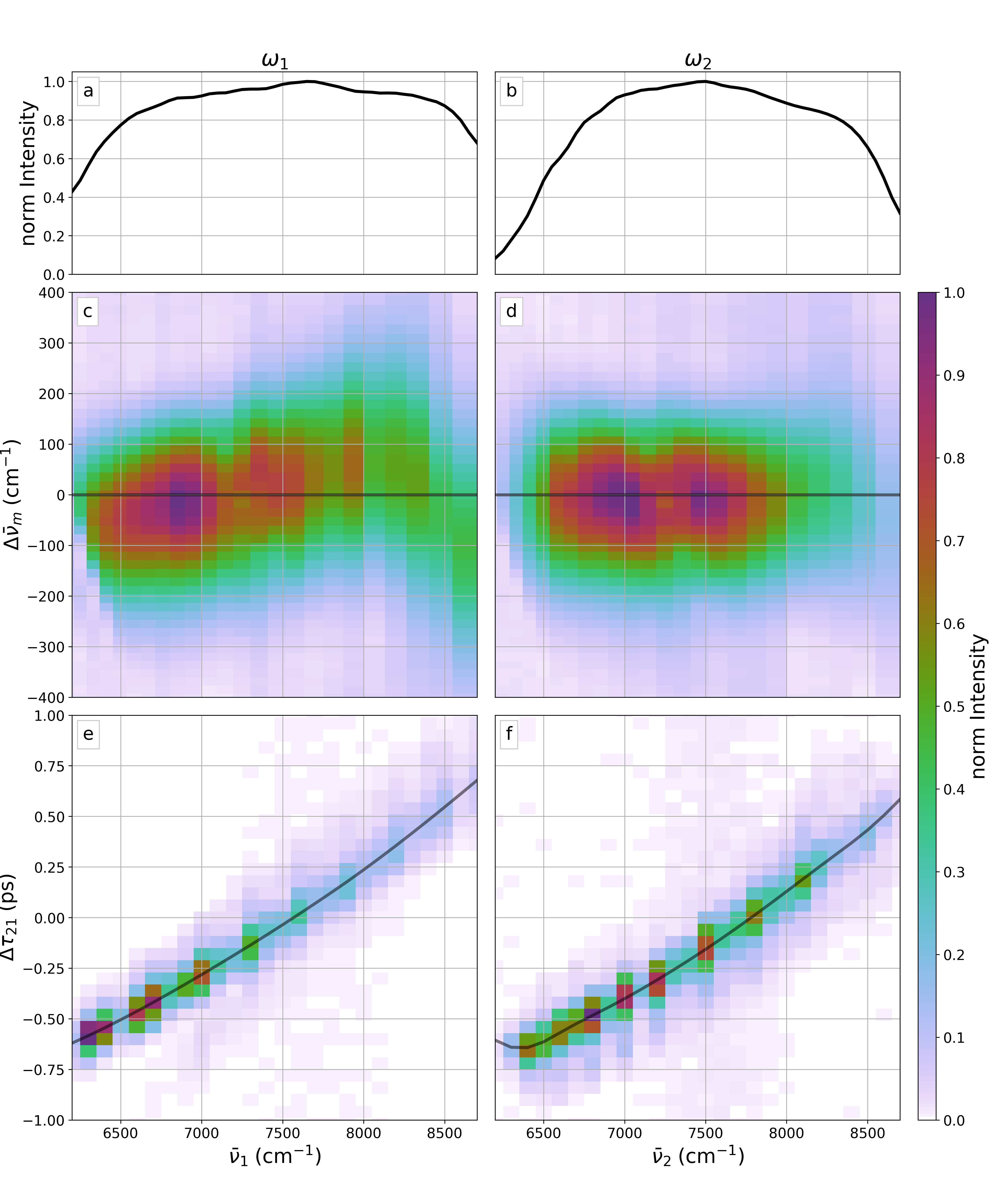}
	\caption{\label{F:OPAs}(Color online) OPA output characterization and correction for different colors of pump light. The left hand column corresponds to $\omega_1$ and the right hand column corresponds to $\omega_2$. Subplots (a) and (b) were acquired by measuring the filtered NIR output of the OPAs with a thermopile [slight smoothing has been applied]. Subplots (c) and (d) were acquired using a monochromator and array detector to spectrally resolve the NIR output of each OPA. Subplots (e) and (f) were acquired by measuring the TSF output of sapphire in transmissive geometry with a PMT and scanning monochromator.}
\end{figure*}

We characterize the OPA outputs using two principle metrics: 
\begin{itemize}
	\item Measuring the output power for each color with a thermopile (Newport, 407A).
	\item Measuring the output spectrum for a given set-point with a home-built InGaAs array detector (Sensor: Hamamatsu, G9494-256D) coupled to a monochromator/spectrograph. 
\end{itemize}
In Fig. \ref{F:OPAs}a-d we show these metrics for both OPAs prior to collection of the data which is presented in the main article. 

We correct for the color-dependent arrival times of incident pulses which we attribute to the dispersion of transmissive optics. 
The corrections that we apply control the arrival times of the driving pulses relative to each-other. 
The data we use to build our corrections are shown in Fig. \ref{F:OPAs}e,f. 
These data were acquired by performing TSF in a transmissive geometry while scanning both delay and set-point frequency for a given OPA with the other OPA set to 7700 cm\textsuperscript{-1}. 
Note how slight periodicities are present along the set-point axis---these are phase-mismatch fringes. 
We splined over these data and then actively offset pulses from each other for \emph{every} pulse color combination. 
We did not take into account the effects of our silicon filter due to it being added to the system after corrections were applied.
\clearpage

\section{Determination of pulse bandwidth}
We determine our approximate pulse bandwidth by taking the data present in Fig. \ref{F:OPAs}d, summing/binning along set-point frequency and then fitting the result to a Gaussian function. 
We find our driving pulses to have a width, on the intensity level, of $\sigma_I = 112\; \mathrm{cm}^{-1}$ which corresponds to an amplitude level width of $\sigma = \sqrt{2} \sigma_I = 160\; \mathrm{cm}^{-1}$. 
\begin{figure}[!htbp]
	\includegraphics[width=.5\textwidth]{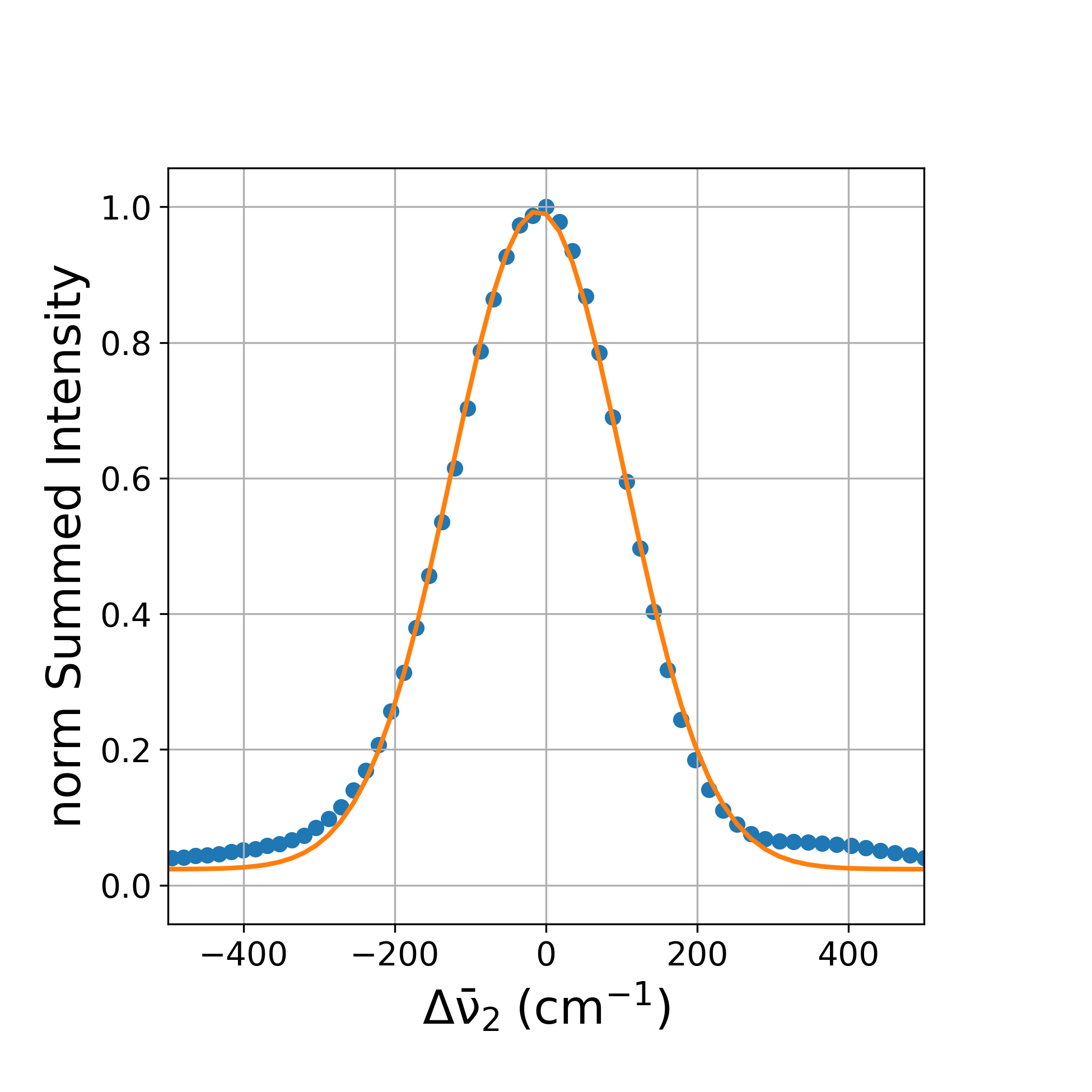}
	\caption{\label{F:OPA_linewidth}(Color online) Determination of pulse bandwidth. Data is blue points while fit is orange line.}
\end{figure}

\bibliography{mybib}

\begin{thebibliography}{42}%
\makeatletter
\providecommand \@ifxundefined [1]{%
 \@ifx{#1\undefined}
}%
\providecommand \@ifnum [1]{%
 \ifnum #1\expandafter \@firstoftwo
 \else \expandafter \@secondoftwo
 \fi
}%
\providecommand \@ifx [1]{%
 \ifx #1\expandafter \@firstoftwo
 \else \expandafter \@secondoftwo
 \fi
}%
\providecommand \natexlab [1]{#1}%
\providecommand \enquote  [1]{``#1''}%
\providecommand \bibnamefont  [1]{#1}%
\providecommand \bibfnamefont [1]{#1}%
\providecommand \citenamefont [1]{#1}%
\providecommand \href@noop [0]{\@secondoftwo}%
\providecommand \href [0]{\begingroup \@sanitize@url \@href}%
\providecommand \@href[1]{\@@startlink{#1}\@@href}%
\providecommand \@@href[1]{\endgroup#1\@@endlink}%
\providecommand \@sanitize@url [0]{\catcode `\\12\catcode `\$12\catcode
  `\&12\catcode `\#12\catcode `\^12\catcode `\_12\catcode `\%12\relax}%
\providecommand \@@startlink[1]{}%
\providecommand \@@endlink[0]{}%
\providecommand \url  [0]{\begingroup\@sanitize@url \@url }%
\providecommand \@url [1]{\endgroup\@href {#1}{\urlprefix }}%
\providecommand \urlprefix  [0]{URL }%
\providecommand \Eprint [0]{\href }%
\providecommand \doibase [0]{http://dx.doi.org/}%
\providecommand \selectlanguage [0]{\@gobble}%
\providecommand \bibinfo  [0]{\@secondoftwo}%
\providecommand \bibfield  [0]{\@secondoftwo}%
\providecommand \translation [1]{[#1]}%
\providecommand \BibitemOpen [0]{}%
\providecommand \bibitemStop [0]{}%
\providecommand \bibitemNoStop [0]{.\EOS\space}%
\providecommand \EOS [0]{\spacefactor3000\relax}%
\providecommand \BibitemShut  [1]{\csname bibitem#1\endcsname}%
\let\auto@bib@innerbib\@empty
\bibitem [{\citenamefont {Boyle}\ \emph
  {et~al.}(2013{\natexlab{a}})\citenamefont {Boyle}, \citenamefont {Pakoulev},\
  and\ \citenamefont {Wright}}]{Boyle_Wright_2013}%
  \BibitemOpen
  \bibfield  {author} {\bibinfo {author} {\bibfnamefont {E.~S.}\ \bibnamefont
  {Boyle}}, \bibinfo {author} {\bibfnamefont {A.~V.}\ \bibnamefont {Pakoulev}},
  \ and\ \bibinfo {author} {\bibfnamefont {J.~C.}\ \bibnamefont {Wright}},\
  }\href {\doibase 10.1021/jp404713x} {\bibfield  {journal} {\bibinfo
  {journal} {The Journal of Physical Chemistry A}\ }\textbf {\bibinfo {volume}
  {117}},\ \bibinfo {pages} {5578} (\bibinfo {year}
  {2013}{\natexlab{a}})}\BibitemShut {NoStop}%
\bibitem [{\citenamefont {Boyle}\ \emph
  {et~al.}(2013{\natexlab{b}})\citenamefont {Boyle}, \citenamefont
  {Neff-Mallon},\ and\ \citenamefont {Wright}}]{Boyle_Wright_2013_001}%
  \BibitemOpen
  \bibfield  {author} {\bibinfo {author} {\bibfnamefont {E.~S.}\ \bibnamefont
  {Boyle}}, \bibinfo {author} {\bibfnamefont {N.~A.}\ \bibnamefont
  {Neff-Mallon}}, \ and\ \bibinfo {author} {\bibfnamefont {J.~C.}\ \bibnamefont
  {Wright}},\ }\href {\doibase 10.1021/jp409377a} {\bibfield  {journal}
  {\bibinfo  {journal} {The Journal of Physical Chemistry A}\ }\textbf
  {\bibinfo {volume} {117}},\ \bibinfo {pages} {12401} (\bibinfo {year}
  {2013}{\natexlab{b}})}\BibitemShut {NoStop}%
\bibitem [{\citenamefont {Boyle}\ \emph {et~al.}(2014)\citenamefont {Boyle},
  \citenamefont {Neff-Mallon}, \citenamefont {Handali},\ and\ \citenamefont
  {Wright}}]{Boyle_Wright_2014}%
  \BibitemOpen
  \bibfield  {author} {\bibinfo {author} {\bibfnamefont {E.~S.}\ \bibnamefont
  {Boyle}}, \bibinfo {author} {\bibfnamefont {N.~A.}\ \bibnamefont
  {Neff-Mallon}}, \bibinfo {author} {\bibfnamefont {J.~D.}\ \bibnamefont
  {Handali}}, \ and\ \bibinfo {author} {\bibfnamefont {J.~C.}\ \bibnamefont
  {Wright}},\ }\href {\doibase 10.1021/jp5018554} {\bibfield  {journal}
  {\bibinfo  {journal} {The Journal of Physical Chemistry A}\ }\textbf
  {\bibinfo {volume} {118}},\ \bibinfo {pages} {3112} (\bibinfo {year}
  {2014})}\BibitemShut {NoStop}%
\bibitem [{\citenamefont {Wright}(2017)}]{Wright_2017}%
  \BibitemOpen
  \bibfield  {author} {\bibinfo {author} {\bibfnamefont {J.~C.}\ \bibnamefont
  {Wright}},\ }\href {\doibase 10.1146/annurev-anchem-061516-045349} {\bibfield
   {journal} {\bibinfo  {journal} {Annual Review of Analytical Chemistry}\
  }\textbf {\bibinfo {volume} {10}},\ \bibinfo {pages} {45} (\bibinfo {year}
  {2017})}\BibitemShut {NoStop}%
\bibitem [{\citenamefont {Armstrong}\ \emph {et~al.}(1962)\citenamefont
  {Armstrong}, \citenamefont {Bloembergen}, \citenamefont {Ducuing},\ and\
  \citenamefont {Pershan}}]{Armstrong_Pershan_1962}%
  \BibitemOpen
  \bibfield  {author} {\bibinfo {author} {\bibfnamefont {J.~A.}\ \bibnamefont
  {Armstrong}}, \bibinfo {author} {\bibfnamefont {N.}~\bibnamefont
  {Bloembergen}}, \bibinfo {author} {\bibfnamefont {J.}~\bibnamefont
  {Ducuing}}, \ and\ \bibinfo {author} {\bibfnamefont {P.~S.}\ \bibnamefont
  {Pershan}},\ }\href {\doibase 10.1103/physrev.127.1918} {\bibfield  {journal}
  {\bibinfo  {journal} {Physical Review}\ }\textbf {\bibinfo {volume} {127}},\
  \bibinfo {pages} {1918} (\bibinfo {year} {1962})}\BibitemShut {NoStop}%
\bibitem [{\citenamefont {Bloembergen}(1965)}]{Bloembergen_1965}%
  \BibitemOpen
  \bibfield  {author} {\bibinfo {author} {\bibfnamefont {N.}~\bibnamefont
  {Bloembergen}},\ }\href@noop {} {\emph {\bibinfo {title} {Nonlinear
  Optics}}}\ (\bibinfo  {publisher} {W. A. Benjamin, Inc.},\ \bibinfo {year}
  {1965})\BibitemShut {NoStop}%
\bibitem [{\citenamefont {Stoker}\ \emph {et~al.}(2005)\citenamefont {Stoker},
  \citenamefont {Becker},\ and\ \citenamefont {Keto}}]{Stoker_Keto_2005}%
  \BibitemOpen
  \bibfield  {author} {\bibinfo {author} {\bibfnamefont {D.}~\bibnamefont
  {Stoker}}, \bibinfo {author} {\bibfnamefont {M.~F.}\ \bibnamefont {Becker}},
  \ and\ \bibinfo {author} {\bibfnamefont {J.~W.}\ \bibnamefont {Keto}},\
  }\href {\doibase 10.1103/physreva.71.061802} {\bibfield  {journal} {\bibinfo
  {journal} {Physical Review A}\ }\textbf {\bibinfo {volume} {71}} (\bibinfo
  {year} {2005}),\ 10.1103/physreva.71.061802}\BibitemShut {NoStop}%
\bibitem [{\citenamefont {Stoker}\ \emph {et~al.}(2006)\citenamefont {Stoker},
  \citenamefont {Baek}, \citenamefont {Wang}, \citenamefont {Kovar},
  \citenamefont {Becker},\ and\ \citenamefont {Keto}}]{Stoker_Keto_2006}%
  \BibitemOpen
  \bibfield  {author} {\bibinfo {author} {\bibfnamefont {D.~S.}\ \bibnamefont
  {Stoker}}, \bibinfo {author} {\bibfnamefont {J.}~\bibnamefont {Baek}},
  \bibinfo {author} {\bibfnamefont {W.}~\bibnamefont {Wang}}, \bibinfo {author}
  {\bibfnamefont {D.}~\bibnamefont {Kovar}}, \bibinfo {author} {\bibfnamefont
  {M.~F.}\ \bibnamefont {Becker}}, \ and\ \bibinfo {author} {\bibfnamefont
  {J.~W.}\ \bibnamefont {Keto}},\ }\href {\doibase 10.1103/physreva.73.053812}
  {\bibfield  {journal} {\bibinfo  {journal} {Physical Review A}\ }\textbf
  {\bibinfo {volume} {73}} (\bibinfo {year} {2006}),\
  10.1103/physreva.73.053812}\BibitemShut {NoStop}%
\bibitem [{\citenamefont {Angerer}\ \emph {et~al.}(1999)\citenamefont
  {Angerer}, \citenamefont {Yang}, \citenamefont {Yodh}, \citenamefont {Khan},\
  and\ \citenamefont {Sun}}]{Angerer_Sun_1999}%
  \BibitemOpen
  \bibfield  {author} {\bibinfo {author} {\bibfnamefont {W.~E.}\ \bibnamefont
  {Angerer}}, \bibinfo {author} {\bibfnamefont {N.}~\bibnamefont {Yang}},
  \bibinfo {author} {\bibfnamefont {A.~G.}\ \bibnamefont {Yodh}}, \bibinfo
  {author} {\bibfnamefont {M.~A.}\ \bibnamefont {Khan}}, \ and\ \bibinfo
  {author} {\bibfnamefont {C.~J.}\ \bibnamefont {Sun}},\ }\href {\doibase
  10.1103/physrevb.59.2932} {\bibfield  {journal} {\bibinfo  {journal}
  {Physical Review B}\ }\textbf {\bibinfo {volume} {59}},\ \bibinfo {pages}
  {2932} (\bibinfo {year} {1999})}\BibitemShut {NoStop}%
\bibitem [{\citenamefont {Boyd}(2008)}]{Boyd_2008}%
  \BibitemOpen
  \bibfield  {author} {\bibinfo {author} {\bibfnamefont {R.~W.}\ \bibnamefont
  {Boyd}},\ }\href@noop {} {\emph {\bibinfo {title} {Nonlinear Optics}}},\
  \bibinfo {edition} {3rd}\ ed.\ (\bibinfo  {publisher} {Academic Press},\
  \bibinfo {year} {2008})\BibitemShut {NoStop}%
\bibitem [{\citenamefont {Tasgal}\ and\ \citenamefont
  {Band}(2004)}]{Tasgal_Band_2004}%
  \BibitemOpen
  \bibfield  {author} {\bibinfo {author} {\bibfnamefont {R.~S.}\ \bibnamefont
  {Tasgal}}\ and\ \bibinfo {author} {\bibfnamefont {Y.~B.}\ \bibnamefont
  {Band}},\ }\href {\doibase 10.1103/physreva.70.053810} {\bibfield  {journal}
  {\bibinfo  {journal} {Physical Review A}\ }\textbf {\bibinfo {volume} {70}}
  (\bibinfo {year} {2004}),\ 10.1103/physreva.70.053810}\BibitemShut {NoStop}%
\bibitem [{Note1()}]{Note1}%
  \BibitemOpen
  \bibinfo {note} {We consider the effects of disregarding $\left . \protect
  \frac {\partial ^2 k}{\partial \omega ^2} \right |_{\omega _0}$ in our Taylor
  expansion by comparing the accrued pulse duration (chirp), $\delta t \approx
  \left . \protect \frac {\partial ^2 k}{\partial ^2 \omega } \right |_{\omega
  _0} \Delta \omega \cdot L$, to the original pulse duration, $\Delta t \approx
  \protect \frac {0.441 \cdot 2}{\Delta \omega }$. For our sample and
  experimental conditions we find $\protect \frac {\delta t}{\Delta t} \approx
  0.1 \%$. Additionally, we may consider the prominence of self-phase
  modulation for our experimental conditions. \protect \rev@citet
  {Siegman_1986} notes that the length scale, $L$, over which self-phase
  modulation is significant goes as $L = \protect \frac {\lambda }{2\pi I
  n_{2I}}$ where $I$ is the intensity at the beam waist and $n_{2I}$ is the
  nonlinear refractive index. Using our experimental parameters and \protect
  \rev@citet {Major_Smith_2004}'s value of $n_{2I} = 3\times 10^{-16} \protect
  \tmspace +\thickmuskip {.2777em} \protect \textrm {cm}^2/\protect \textrm
  {W}$ for $\lambda =1300 \protect \tmspace +\thickmuskip {.2777em} \protect
  \textrm {nm}$, we find that $L\approx 0.3 \protect \tmspace +\thickmuskip
  {.2777em} \protect \textrm {cm}$, which is an order of magnitude longer than
  our sample. Given both of these calculations, second-order effects are much
  smaller than first-order effects given our pulse bandwidth and sample length.
  We think of the intensity of driving fields at which our work is accomplished
  as being sufficient to see generated third-order response against black, but
  not sufficient to observe third-order effects (self-phase modulation) in the
  driving fields.}\BibitemShut {Stop}%
\bibitem [{\citenamefont {Maker}\ and\ \citenamefont
  {Terhune}(1965)}]{Maker_Terhune_1965}%
  \BibitemOpen
  \bibfield  {author} {\bibinfo {author} {\bibfnamefont {P.~D.}\ \bibnamefont
  {Maker}}\ and\ \bibinfo {author} {\bibfnamefont {R.~W.}\ \bibnamefont
  {Terhune}},\ }\href {\doibase 10.1103/physrev.137.a801} {\bibfield  {journal}
  {\bibinfo  {journal} {Physical Review}\ }\textbf {\bibinfo {volume} {137}},\
  \bibinfo {pages} {A801} (\bibinfo {year} {1965})}\BibitemShut {NoStop}%
\bibitem [{\citenamefont {Kohler}\ \emph {et~al.}(2017)\citenamefont {Kohler},
  \citenamefont {Thompson},\ and\ \citenamefont {Wright}}]{Kohler_Wright_2017}%
  \BibitemOpen
  \bibfield  {author} {\bibinfo {author} {\bibfnamefont {D.~D.}\ \bibnamefont
  {Kohler}}, \bibinfo {author} {\bibfnamefont {B.~J.}\ \bibnamefont
  {Thompson}}, \ and\ \bibinfo {author} {\bibfnamefont {J.~C.}\ \bibnamefont
  {Wright}},\ }\href {\doibase 10.1063/1.4986069} {\bibfield  {journal}
  {\bibinfo  {journal} {The Journal of Chemical Physics}\ }\textbf {\bibinfo
  {volume} {147}},\ \bibinfo {pages} {084202} (\bibinfo {year}
  {2017})}\BibitemShut {NoStop}%
\bibitem [{Note2()}]{Note2}%
  \BibitemOpen
  \bibinfo {note} {We use the well-known relation $\DOTSI \intop \ilimits@
  _{-\infty }^{\infty } \protect \qopname \relax o{exp}{\left [-\left
  (ax^2+bx\right )\right ]}\protect \textrm {d}x = \protect \sqrt {\protect
  \frac {\pi }{a}}\protect \qopname \relax o{exp}{\left [\protect \frac
  {b^2}{4a}\right ]}$ to integrate Eqn. \ref {E:polar}.}\BibitemShut {Stop}%
\bibitem [{\citenamefont {Shen}(1984)}]{Shen_1984}%
  \BibitemOpen
  \bibfield  {author} {\bibinfo {author} {\bibfnamefont {Y.~R.}\ \bibnamefont
  {Shen}},\ }\href@noop {} {\emph {\bibinfo {title} {The Principles of
  Nonlinear Optics}}},\ \bibinfo {edition} {1st}\ ed.\ (\bibinfo  {publisher}
  {John Wiley \& Sons},\ \bibinfo {year} {1984})\BibitemShut {NoStop}%
\bibitem [{\citenamefont {Maker}\ \emph {et~al.}(1962)\citenamefont {Maker},
  \citenamefont {Terhune}, \citenamefont {Nisenoff},\ and\ \citenamefont
  {Savage}}]{Maker_Savage_1962}%
  \BibitemOpen
  \bibfield  {author} {\bibinfo {author} {\bibfnamefont {P.~D.}\ \bibnamefont
  {Maker}}, \bibinfo {author} {\bibfnamefont {R.~W.}\ \bibnamefont {Terhune}},
  \bibinfo {author} {\bibfnamefont {M.}~\bibnamefont {Nisenoff}}, \ and\
  \bibinfo {author} {\bibfnamefont {C.~M.}\ \bibnamefont {Savage}},\ }\href
  {\doibase 10.1103/physrevlett.8.21} {\bibfield  {journal} {\bibinfo
  {journal} {Physical Review Letters}\ }\textbf {\bibinfo {volume} {8}},\
  \bibinfo {pages} {21} (\bibinfo {year} {1962})}\BibitemShut {NoStop}%
\bibitem [{\citenamefont {Mlejnek}\ \emph {et~al.}(1999)\citenamefont
  {Mlejnek}, \citenamefont {Wright}, \citenamefont {Moloney},\ and\
  \citenamefont {Bloembergen}}]{Mlejnek_Bloembergen_1999}%
  \BibitemOpen
  \bibfield  {author} {\bibinfo {author} {\bibfnamefont {M.}~\bibnamefont
  {Mlejnek}}, \bibinfo {author} {\bibfnamefont {E.~M.}\ \bibnamefont {Wright}},
  \bibinfo {author} {\bibfnamefont {J.~V.}\ \bibnamefont {Moloney}}, \ and\
  \bibinfo {author} {\bibfnamefont {N.}~\bibnamefont {Bloembergen}},\ }\href
  {\doibase 10.1103/physrevlett.83.2934} {\bibfield  {journal} {\bibinfo
  {journal} {Physical Review Letters}\ }\textbf {\bibinfo {volume} {83}},\
  \bibinfo {pages} {2934} (\bibinfo {year} {1999})}\BibitemShut {NoStop}%
\bibitem [{\citenamefont {Trabs}\ \emph {et~al.}(2015)\citenamefont {Trabs},
  \citenamefont {Noack}, \citenamefont {Aleksandrovsky}, \citenamefont
  {Zaitsev}, \citenamefont {Radionov},\ and\ \citenamefont
  {Petrov}}]{Trabs_Petrov_2015}%
  \BibitemOpen
  \bibfield  {author} {\bibinfo {author} {\bibfnamefont {P.}~\bibnamefont
  {Trabs}}, \bibinfo {author} {\bibfnamefont {F.}~\bibnamefont {Noack}},
  \bibinfo {author} {\bibfnamefont {A.~S.}\ \bibnamefont {Aleksandrovsky}},
  \bibinfo {author} {\bibfnamefont {A.~I.}\ \bibnamefont {Zaitsev}}, \bibinfo
  {author} {\bibfnamefont {N.~V.}\ \bibnamefont {Radionov}}, \ and\ \bibinfo
  {author} {\bibfnamefont {V.}~\bibnamefont {Petrov}},\ }\href {\doibase
  10.1364/oe.23.010091} {\bibfield  {journal} {\bibinfo  {journal} {Optics
  Express}\ }\textbf {\bibinfo {volume} {23}},\ \bibinfo {pages} {10091}
  (\bibinfo {year} {2015})}\BibitemShut {NoStop}%
\bibitem [{\citenamefont {Rassoul}\ \emph {et~al.}(1997)\citenamefont
  {Rassoul}, \citenamefont {Ivanov}, \citenamefont {Freysz}, \citenamefont
  {Ducasse},\ and\ \citenamefont {Hache}}]{Rassoul_Hache_1997}%
  \BibitemOpen
  \bibfield  {author} {\bibinfo {author} {\bibfnamefont {R.~M.}\ \bibnamefont
  {Rassoul}}, \bibinfo {author} {\bibfnamefont {A.}~\bibnamefont {Ivanov}},
  \bibinfo {author} {\bibfnamefont {E.}~\bibnamefont {Freysz}}, \bibinfo
  {author} {\bibfnamefont {A.}~\bibnamefont {Ducasse}}, \ and\ \bibinfo
  {author} {\bibfnamefont {F.}~\bibnamefont {Hache}},\ }\href {\doibase
  10.1364/ol.22.000268} {\bibfield  {journal} {\bibinfo  {journal} {Optics
  Letters}\ }\textbf {\bibinfo {volume} {22}},\ \bibinfo {pages} {268}
  (\bibinfo {year} {1997})}\BibitemShut {NoStop}%
\bibitem [{\citenamefont {Glenn}(1969)}]{Glenn_1969}%
  \BibitemOpen
  \bibfield  {author} {\bibinfo {author} {\bibfnamefont {W.}~\bibnamefont
  {Glenn}},\ }\href {\doibase 10.1109/jqe.1969.1081948} {\bibfield  {journal}
  {\bibinfo  {journal} {{IEEE} Journal of Quantum Electronics}\ }\textbf
  {\bibinfo {volume} {5}},\ \bibinfo {pages} {284} (\bibinfo {year}
  {1969})}\BibitemShut {NoStop}%
\bibitem [{\citenamefont {Sidick}\ \emph {et~al.}(1995)\citenamefont {Sidick},
  \citenamefont {Knoesen},\ and\ \citenamefont {Dienes}}]{Sidick_Dienes_1995}%
  \BibitemOpen
  \bibfield  {author} {\bibinfo {author} {\bibfnamefont {E.}~\bibnamefont
  {Sidick}}, \bibinfo {author} {\bibfnamefont {A.}~\bibnamefont {Knoesen}}, \
  and\ \bibinfo {author} {\bibfnamefont {A.}~\bibnamefont {Dienes}},\ }\href
  {\doibase 10.1364/josab.12.001704} {\bibfield  {journal} {\bibinfo  {journal}
  {Journal of the Optical Society of America B}\ }\textbf {\bibinfo {volume}
  {12}},\ \bibinfo {pages} {1704} (\bibinfo {year} {1995})}\BibitemShut
  {NoStop}%
\bibitem [{\citenamefont {Diels}\ and\ \citenamefont
  {Rudolph}(1996)}]{Diels_Rudolph_1996}%
  \BibitemOpen
  \bibfield  {author} {\bibinfo {author} {\bibfnamefont {J.-C.}\ \bibnamefont
  {Diels}}\ and\ \bibinfo {author} {\bibfnamefont {W.}~\bibnamefont
  {Rudolph}},\ }\href@noop {} {\emph {\bibinfo {title} {Ultrashort Laser Pulse
  Phenomena: Fundamentals, Techniques, and Applications on a Femtosecond Time
  Scale}}},\ \bibinfo {edition} {1st}\ ed.\ (\bibinfo  {publisher} {Academic
  Press},\ \bibinfo {year} {1996})\BibitemShut {NoStop}%
\bibitem [{\citenamefont {Malitson}(1962)}]{Malitson_1962}%
  \BibitemOpen
  \bibfield  {author} {\bibinfo {author} {\bibfnamefont {I.~H.}\ \bibnamefont
  {Malitson}},\ }\href {\doibase 10.1364/josa.52.001377} {\bibfield  {journal}
  {\bibinfo  {journal} {Journal of the Optical Society of America}\ }\textbf
  {\bibinfo {volume} {52}},\ \bibinfo {pages} {1377} (\bibinfo {year}
  {1962})}\BibitemShut {NoStop}%
\bibitem [{\citenamefont {Manassah}(1988)}]{Manassah_1988}%
  \BibitemOpen
  \bibfield  {author} {\bibinfo {author} {\bibfnamefont {J.~T.}\ \bibnamefont
  {Manassah}},\ }\href {\doibase 10.1364/ao.27.004365} {\bibfield  {journal}
  {\bibinfo  {journal} {Applied Optics}\ }\textbf {\bibinfo {volume} {27}},\
  \bibinfo {pages} {4365} (\bibinfo {year} {1988})}\BibitemShut {NoStop}%
\bibitem [{\citenamefont {Noordam}\ \emph {et~al.}(1990)\citenamefont
  {Noordam}, \citenamefont {Bakker}, \citenamefont {de~Boer},\ and\
  \citenamefont {van Linden van~den
  Heuvell}}]{Noordam_vanLindenvandenHeuvell_1990}%
  \BibitemOpen
  \bibfield  {author} {\bibinfo {author} {\bibfnamefont {L.~D.}\ \bibnamefont
  {Noordam}}, \bibinfo {author} {\bibfnamefont {H.~J.}\ \bibnamefont {Bakker}},
  \bibinfo {author} {\bibfnamefont {M.~P.}\ \bibnamefont {de~Boer}}, \ and\
  \bibinfo {author} {\bibfnamefont {H.~B.}\ \bibnamefont {van Linden van~den
  Heuvell}},\ }\href {\doibase 10.1364/ol.15.001464} {\bibfield  {journal}
  {\bibinfo  {journal} {Optics Letters}\ }\textbf {\bibinfo {volume} {15}},\
  \bibinfo {pages} {1464} (\bibinfo {year} {1990})}\BibitemShut {NoStop}%
\bibitem [{\citenamefont {van Rossum}\ \emph {et~al.}(01  )\citenamefont {van
  Rossum} \emph {et~al.}}]{vanRossum_2001}%
  \BibitemOpen
  \bibfield  {author} {\bibinfo {author} {\bibfnamefont {G.}~\bibnamefont {van
  Rossum}} \emph {et~al.},\ }\href {http://www.python.org/} {\enquote {\bibinfo
  {title} {{Python}},}\ } (\bibinfo {year} {2001--}),\ \bibinfo {note}
  {[Online; accessed 2017-09-28]}\BibitemShut {NoStop}%
\bibitem [{\citenamefont {Jones}\ \emph {et~al.}(01  )\citenamefont {Jones},
  \citenamefont {Oliphant}, \citenamefont {Peterson} \emph
  {et~al.}}]{Jones_2001}%
  \BibitemOpen
  \bibfield  {author} {\bibinfo {author} {\bibfnamefont {E.}~\bibnamefont
  {Jones}}, \bibinfo {author} {\bibfnamefont {T.}~\bibnamefont {Oliphant}},
  \bibinfo {author} {\bibfnamefont {P.}~\bibnamefont {Peterson}},  \emph
  {et~al.},\ }\href {http://www.scipy.org/} {\enquote {\bibinfo {title}
  {{SciPy}: Open source scientific tools for {Python}},}\ } (\bibinfo {year}
  {2001--}),\ \bibinfo {note} {[Online; accessed 2017-09-28]}\BibitemShut
  {NoStop}%
\bibitem [{\citenamefont {van~der Walt}\ \emph {et~al.}(2011)\citenamefont
  {van~der Walt}, \citenamefont {Colbert},\ and\ \citenamefont
  {Varoquaux}}]{vanderWalt_Varoquaux_2011}%
  \BibitemOpen
  \bibfield  {author} {\bibinfo {author} {\bibfnamefont {S.}~\bibnamefont
  {van~der Walt}}, \bibinfo {author} {\bibfnamefont {S.~C.}\ \bibnamefont
  {Colbert}}, \ and\ \bibinfo {author} {\bibfnamefont {G.}~\bibnamefont
  {Varoquaux}},\ }\href {\doibase 10.1109/mcse.2011.37} {\bibfield  {journal}
  {\bibinfo  {journal} {Computing in Science {\&} Engineering}\ }\textbf
  {\bibinfo {volume} {13}},\ \bibinfo {pages} {22} (\bibinfo {year}
  {2011})}\BibitemShut {NoStop}%
\bibitem [{\citenamefont {Hunter}(2007)}]{Hunter_2007}%
  \BibitemOpen
  \bibfield  {author} {\bibinfo {author} {\bibfnamefont {J.~D.}\ \bibnamefont
  {Hunter}},\ }\href {\doibase 10.1109/mcse.2007.55} {\bibfield  {journal}
  {\bibinfo  {journal} {Computing in Science {\&} Engineering}\ }\textbf
  {\bibinfo {volume} {9}},\ \bibinfo {pages} {90} (\bibinfo {year}
  {2007})}\BibitemShut {NoStop}%
\bibitem [{Note3()}]{Note3}%
  \BibitemOpen
  \bibinfo {note} {The spectra shown in Fig. \ref {F:w1w2}b was acquired with
  our monochromator/grating in ``0th order mode'' which effectively passed all
  colors to the detector as if it were a lossy mirror.}\BibitemShut {Stop}%
\bibitem [{\citenamefont {Murdoch}\ \emph {et~al.}(2000)\citenamefont
  {Murdoch}, \citenamefont {Thompson}, \citenamefont {Meyer},\ and\
  \citenamefont {Wright}}]{Murdoch_Wright_2000}%
  \BibitemOpen
  \bibfield  {author} {\bibinfo {author} {\bibfnamefont {K.~M.}\ \bibnamefont
  {Murdoch}}, \bibinfo {author} {\bibfnamefont {D.~E.}\ \bibnamefont
  {Thompson}}, \bibinfo {author} {\bibfnamefont {K.~A.}\ \bibnamefont {Meyer}},
  \ and\ \bibinfo {author} {\bibfnamefont {J.~C.}\ \bibnamefont {Wright}},\
  }\href {\doibase 10.1366/0003702001948411} {\bibfield  {journal} {\bibinfo
  {journal} {Applied Spectroscopy}\ }\textbf {\bibinfo {volume} {54}},\
  \bibinfo {pages} {1495} (\bibinfo {year} {2000})}\BibitemShut {NoStop}%
\bibitem [{\citenamefont {Volkmer}\ \emph {et~al.}(2001)\citenamefont
  {Volkmer}, \citenamefont {Cheng},\ and\ \citenamefont
  {Xie}}]{Volkmer_Xie_2001}%
  \BibitemOpen
  \bibfield  {author} {\bibinfo {author} {\bibfnamefont {A.}~\bibnamefont
  {Volkmer}}, \bibinfo {author} {\bibfnamefont {J.-X.}\ \bibnamefont {Cheng}},
  \ and\ \bibinfo {author} {\bibfnamefont {X.~S.}\ \bibnamefont {Xie}},\ }\href
  {\doibase 10.1103/physrevlett.87.023901} {\bibfield  {journal} {\bibinfo
  {journal} {Physical Review Letters}\ }\textbf {\bibinfo {volume} {87}}
  (\bibinfo {year} {2001}),\ 10.1103/physrevlett.87.023901}\BibitemShut
  {NoStop}%
\bibitem [{\citenamefont {Czech}\ \emph {et~al.}(2015)\citenamefont {Czech},
  \citenamefont {Thompson}, \citenamefont {Kain}, \citenamefont {Ding},
  \citenamefont {Shearer}, \citenamefont {Hamers}, \citenamefont {Jin},\ and\
  \citenamefont {Wright}}]{Czech_Wright_2015}%
  \BibitemOpen
  \bibfield  {author} {\bibinfo {author} {\bibfnamefont {K.~J.}\ \bibnamefont
  {Czech}}, \bibinfo {author} {\bibfnamefont {B.~J.}\ \bibnamefont {Thompson}},
  \bibinfo {author} {\bibfnamefont {S.}~\bibnamefont {Kain}}, \bibinfo {author}
  {\bibfnamefont {Q.}~\bibnamefont {Ding}}, \bibinfo {author} {\bibfnamefont
  {M.~J.}\ \bibnamefont {Shearer}}, \bibinfo {author} {\bibfnamefont {R.~J.}\
  \bibnamefont {Hamers}}, \bibinfo {author} {\bibfnamefont {S.}~\bibnamefont
  {Jin}}, \ and\ \bibinfo {author} {\bibfnamefont {J.~C.}\ \bibnamefont
  {Wright}},\ }\href {\doibase 10.1021/acsnano.5b05198} {\bibfield  {journal}
  {\bibinfo  {journal} {{ACS} Nano}\ }\textbf {\bibinfo {volume} {9}},\
  \bibinfo {pages} {12146} (\bibinfo {year} {2015})}\BibitemShut {NoStop}%
\bibitem [{\citenamefont {Wang}\ and\ \citenamefont
  {Baardsen}(1969)}]{Wang_Baardsen_1969}%
  \BibitemOpen
  \bibfield  {author} {\bibinfo {author} {\bibfnamefont {C.~C.}\ \bibnamefont
  {Wang}}\ and\ \bibinfo {author} {\bibfnamefont {E.~L.}\ \bibnamefont
  {Baardsen}},\ }\href {\doibase 10.1103/physrev.185.1079} {\bibfield
  {journal} {\bibinfo  {journal} {Physical Review}\ }\textbf {\bibinfo {volume}
  {185}},\ \bibinfo {pages} {1079} (\bibinfo {year} {1969})}\BibitemShut
  {NoStop}%
\bibitem [{\citenamefont {Bey}\ \emph {et~al.}(1968)\citenamefont {Bey},
  \citenamefont {Guiliani},\ and\ \citenamefont {Rabin}}]{Bey_Rabin_1968}%
  \BibitemOpen
  \bibfield  {author} {\bibinfo {author} {\bibfnamefont {P.}~\bibnamefont
  {Bey}}, \bibinfo {author} {\bibfnamefont {J.}~\bibnamefont {Guiliani}}, \
  and\ \bibinfo {author} {\bibfnamefont {H.}~\bibnamefont {Rabin}},\ }\href
  {\doibase 10.1016/0375-9601(68)90403-9} {\bibfield  {journal} {\bibinfo
  {journal} {Physics Letters A}\ }\textbf {\bibinfo {volume} {28}},\ \bibinfo
  {pages} {89} (\bibinfo {year} {1968})}\BibitemShut {NoStop}%
\bibitem [{\citenamefont {Crisp}(1970)}]{Crisp_1970}%
  \BibitemOpen
  \bibfield  {author} {\bibinfo {author} {\bibfnamefont {M.~D.}\ \bibnamefont
  {Crisp}},\ }\href {\doibase 10.1103/physreva.1.1604} {\bibfield  {journal}
  {\bibinfo  {journal} {Physical Review A}\ }\textbf {\bibinfo {volume} {1}},\
  \bibinfo {pages} {1604} (\bibinfo {year} {1970})}\BibitemShut {NoStop}%
\bibitem [{\citenamefont {Carlson}\ and\ \citenamefont
  {Wright}(1989)}]{Carlson_Wright_1989}%
  \BibitemOpen
  \bibfield  {author} {\bibinfo {author} {\bibfnamefont {R.~J.}\ \bibnamefont
  {Carlson}}\ and\ \bibinfo {author} {\bibfnamefont {J.~C.}\ \bibnamefont
  {Wright}},\ }\href {\doibase 10.1366/0003702894203408} {\bibfield  {journal}
  {\bibinfo  {journal} {Applied Spectroscopy}\ }\textbf {\bibinfo {volume}
  {43}},\ \bibinfo {pages} {1195} (\bibinfo {year} {1989})}\BibitemShut
  {NoStop}%
\bibitem [{\citenamefont {Kornau}\ \emph {et~al.}(2011)\citenamefont {Kornau},
  \citenamefont {Rickard}, \citenamefont {Mathew}, \citenamefont {Pakoulev},\
  and\ \citenamefont {Wright}}]{Kornau_Wright_2011}%
  \BibitemOpen
  \bibfield  {author} {\bibinfo {author} {\bibfnamefont {K.~M.}\ \bibnamefont
  {Kornau}}, \bibinfo {author} {\bibfnamefont {M.~A.}\ \bibnamefont {Rickard}},
  \bibinfo {author} {\bibfnamefont {N.~A.}\ \bibnamefont {Mathew}}, \bibinfo
  {author} {\bibfnamefont {A.~V.}\ \bibnamefont {Pakoulev}}, \ and\ \bibinfo
  {author} {\bibfnamefont {J.~C.}\ \bibnamefont {Wright}},\ }\href {\doibase
  10.1021/jp1104856} {\bibfield  {journal} {\bibinfo  {journal} {The Journal of
  Physical Chemistry A}\ }\textbf {\bibinfo {volume} {115}},\ \bibinfo {pages}
  {4054} (\bibinfo {year} {2011})}\BibitemShut {NoStop}%
\bibitem [{\citenamefont {Kinrot}\ and\ \citenamefont
  {Prior}(1994)}]{Kinrot_Prior_1994}%
  \BibitemOpen
  \bibfield  {author} {\bibinfo {author} {\bibfnamefont {O.}~\bibnamefont
  {Kinrot}}\ and\ \bibinfo {author} {\bibfnamefont {Y.}~\bibnamefont {Prior}},\
  }\href {\doibase 10.1103/physreva.50.r1999} {\bibfield  {journal} {\bibinfo
  {journal} {Physical Review A}\ }\textbf {\bibinfo {volume} {50}},\ \bibinfo
  {pages} {R1999} (\bibinfo {year} {1994})}\BibitemShut {NoStop}%
\bibitem [{\citenamefont {Siegman}(1986)}]{Siegman_1986}%
  \BibitemOpen
  \bibfield  {author} {\bibinfo {author} {\bibfnamefont {A.~E.}\ \bibnamefont
  {Siegman}},\ }\href@noop {} {\emph {\bibinfo {title} {Lasers}}},\ \bibinfo
  {edition} {1st}\ ed.\ (\bibinfo  {publisher} {University Science Books},\
  \bibinfo {year} {1986})\BibitemShut {NoStop}%
\bibitem [{\citenamefont {Major}\ \emph {et~al.}(2004)\citenamefont {Major},
  \citenamefont {Yoshino}, \citenamefont {Nikolakakos}, \citenamefont
  {Aitchison},\ and\ \citenamefont {Smith}}]{Major_Smith_2004}%
  \BibitemOpen
  \bibfield  {author} {\bibinfo {author} {\bibfnamefont {A.}~\bibnamefont
  {Major}}, \bibinfo {author} {\bibfnamefont {F.}~\bibnamefont {Yoshino}},
  \bibinfo {author} {\bibfnamefont {I.}~\bibnamefont {Nikolakakos}}, \bibinfo
  {author} {\bibfnamefont {J.~S.}\ \bibnamefont {Aitchison}}, \ and\ \bibinfo
  {author} {\bibfnamefont {P.~W.~E.}\ \bibnamefont {Smith}},\ }\href {\doibase
  10.1364/ol.29.000602} {\bibfield  {journal} {\bibinfo  {journal} {Optics
  Letters}\ }\textbf {\bibinfo {volume} {29}},\ \bibinfo {pages} {602}
  (\bibinfo {year} {2004})}\BibitemShut {NoStop}%
\end{thebibliography}%

\end{document}